\documentclass[prl,aps,twocolumn,showpacs,amsmath,amssymb,superscriptaddress]{revtex4-1}
\usepackage{graphicx}
\usepackage{dcolumn}
\usepackage{bm}
\usepackage{subfigure}
\usepackage{setspace}
\usepackage{psfrag}
\usepackage{feynmp}
\newcommand{\mbf}[1]{\boldsymbol{\mathbf{#1}}}

\newcommand{\be}{\begin{equation}}
\newcommand{\ee}{\end{equation}}
\newcommand{\ba}{\begin{array}}

\newcommand{\ea}{\end{array}}
\newcommand{\bqa}{\begin{eqnarray}}
\newcommand{\eqa}{\end{eqnarray}}

\newcommand{\bra}[1]{\ensuremath{\langle #1 |}}
\newcommand{\ket}[1]{\ensuremath{| #1 \rangle}}

\usepackage{bm}
\begin{document}
\title{Transition of a mesoscopic bosonic gas into a Bose-Einstein condensate}
\author{Alexej Schelle}
\affiliation{
Helmholtz Zentrum M\"{u}nchen, Ingolst\"{a}dter Landstrasse 1, D-85764 Neuherberg (present working adress)}
\date{\today}
\begin{abstract}
The condensate number distribution during the transition of a dilute, weakly interacting gas of $N=200$ bosonic atoms 
into a Bose-Einstein condensate is modeled within number conserving master equation theory of Bose-Einstein condensation. 
Initial strong quantum fluctuations occuring during the exponential cycle of condensate growth 
reduce in a subsequent saturation stage, before the Bose gas finally relaxes towards the Gibbs-Boltzmann equilibrium.
\end{abstract}


\maketitle
\section{I. Introduction}
Bose-Einstein condensates have turned into exquisite tools to study fundamental 
quantum phenomena on the micrometer scale, and a vast range of different physical scenarios 
have been realized experimentally with ultracold matter in the last decade, 
confirming the fundamental importance and the broad applicational scope
of Bose-Einstein condensation~\cite{condensation}. 
A microscopic, quantum \textit{dynamical} description of the gas' $N$-body state {during} 
the buildup of the condensed phase after a sudden switch of the gas temperature below the critical 
temperature expected for Bose-Einstein condensation, however, is still one of the most 
striking theoretical topics of ultracold matter physics up to date.

How can we model the quantum many-body dynamics 
during the transition of a dilute, weakly interacting gas of bosonic 
atoms into a Bose-Einstein condensate and link the resulting equilibrium statistics 
of the condensed phase to the roots of Bose statistics, i.e. to the statistics of 
quasi-ideal Bose gases, which was introduced almost one century 
ago by Bose and Einstein~\cite{Einstein}? 
And under which conditions is the equilibrium statistics of a dilute and weakly interacting 
Bose-Einstein condensate of fixed particle number 
uniquely determined by the statistics of an ideal Gibbs-Boltzmann thermal gas  -- 
lacking any hysteresis on the condensate formation process~\cite{Demler}?

Since an exact numerical solution to the full $N$-body problem is out of the range of today's supercomputing facilities already for rather 
small atomic samples with a few hundreds of atoms, these questions reduce to finding appropriate and numerically accessible effective 
equations for the quantum many-body dynamics of the Bose-Einstein phase transition. 
So far, extensive pioneering works~\cite{QKT, Stoof, Walser1} (and references cited therein) 
have led to accurate dynamic equations describing the evolution of 
the \textit{average} condensate population for atomic gases with a few thousands to a few millions of atoms: 
the onset of Bose-Einstein condensation is marked by a spontaneously insetting exponential growth of the average ground state population, 
followed by a slow, subsequent saturation stage in which the condensate fraction converges 
towards an equilibrium value after suddenly cooling~\cite{Esslinger} the gas below its critical temperature.

In this Article, the relaxation dynamics of the entire state of a mesoscopic Bose 
gas during the transition of exactly $N=200$ weakly interacting bosonic atoms into a 
Bose-Einstein condensate is reported for the first time within number-conserving master equation theory of 
Bose-Einstein condensation \cite{TBE, TBE1}. 
In particular, focus is put on following the dynamics of the condensate number distribution 
during the Bose-Einstein condensation process. Detailed theoretical knowledge on the \textit{statistics and dynamics} of 
mesoscopic ($N\sim200-1000$ atoms), weakly interacting 
quantum gases is not only of importance for state-of-the-art 
experiments~\cite{Obe1}, but moreover contrasts fundamental postulates of classical, 
statistical mechanics to interacting, quantum degenerate many particle systems of finite size.
The latter aspect arises from the neglect of number and energy fluctuations in standard thermodynamics, 
which is a reasonable assumption for classical, non-interacting gases 
in the thermodynamic limit (with particle numbers as large as 
Avogadro's number, $N\sim10^{23}$), whereas a vanishing impact of such 
quantum effects in the quantum degenerate limit is a priori no longer irrefragable.

\section{II. Master equation theory}
\begin{figure}[t]
\begin{center}
\centerline{\includegraphics[width=5.8cm,height=4.0cm,angle = 0.0]{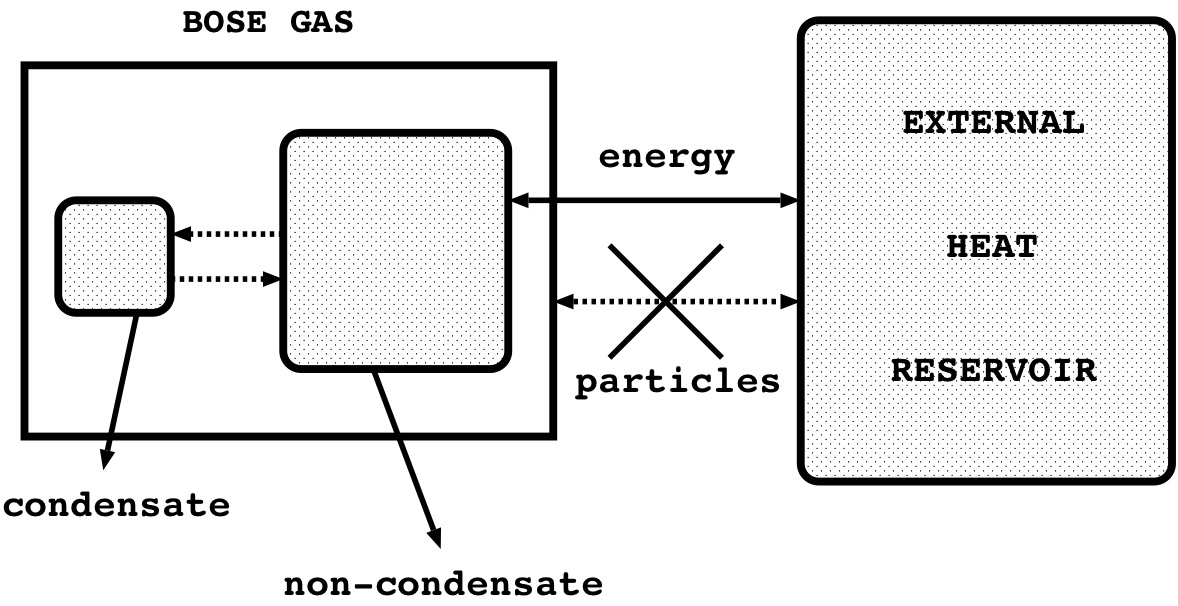}}
\caption{(color online) Schematics of microsopic many particle dynamics. 
The total number of atoms in the Bose gas is fixed to $N$.
Thermal equilibration \textit{within} the non-condensate 
is modeled by a heat reservoir which is at fixed temperature $T$. 
After suddendly cooling the gas, condensate and non-condensate undergo particle exchange due to the atomic two-body collisions until 
a steady state is reached.}
\label{scheme}
\end{center}
\end{figure} 

In order to model Bose-Einstein condensation for small atomic gases, it is sufficient to numerically monitor 
the condensate and non-condensate number distribution during condensation, as will be shown in the following. For this purpose, we derive 
a master equation for said number distributions assuming a 3-dimensional harmonic trapping potential, 
with $p_N(N_{\vec{0}},t)=p_N(N-N_{\vec{0}},t)$ the condensate number distribution and $N_{\vec{0}}$ the number of condensed atoms, 
given a gas of $N$ particles therein. The particle number $N$ may be on the order of a few hundreds 
to a few millions of atoms, but is kept fixed during the final cycle of the condensation process. The derivation of the master equation is lengthy and relies on several 
approximations~\cite{TBE, TBE1}, which can be summarized in the following way. 
We consider condensation onto one single-particle 
(system) mode $\ket{\Psi_{\vec{0}}}$, the condensate mode,
whereas all orthogonal, non-condensate single-particle modes, $\lbrace\ket{\Psi_{\vec{\mbf{k}}}},\vec{\mbf{k}}=(k_x,k_y,k_z)\rbrace$, 
are assumed to be lowly occupied (environment). 
Next, we employ the separation of time scales in dilute atomic gases~\cite{Walser1, formation}. 
The relaxation dynamics in the high energetic, 
non-condensate part of the gas takes place on a time scale of two body collisions, 
$\tau_{{\rm collision}}\sim$ ms, and is thus much faster than the time scale of 
condensate formation, $\tau_0\sim$ sec. 
Assuming this Markovian dynamics scenario which is basically justified by the rapid non-condensate thermalization, we formally map 
each Fock subspace of $(N-N_{\vec{0}})$ non-condensate particles onto a thermal mixture to model the rapid non-condensate 
rethermalization, and to care for particle number conservation. In particular, we replace interactions in the non-condensate thermal bath 
by formally coupling the non-condensate to an external heat bath of fixed temperature $T$. 
Finally, we use the Fock number representation, $\ket{N_{\vec{0}},\lbrace N_{k}\rbrace}$, with $N_{\vec{0}}$, the number of condensate particles, 
and with $N_{\perp}=\sum_{\vec{k}\ne0}N_{\vec{k}}$, the number of non-condensate particles.
The thermodynamical scheme of our description is depicted in fig.~\ref{scheme}.

Since the particle number is conserved, the above ingredients imply that the state of the Bose gas 
in Fock representation of the gas obeys merely classical correlations originating from particle number conservation. A 
Born-Markov ansatz~\cite{Tan} generalized for an $N$-body state of a fixed particle number is thus given by: 

\begin{equation}
\hat{\rho}^{(N)}(t)=\sum_{N_{\vec{\mbf{0}}}=0}^{N}p_{N}(N_{\vec{\mbf{0}}},t)\ket{N_{\vec{\mbf{0}}}}
\bra{N_{\vec{\mbf{0}}}}\otimes\hat{\rho}_\perp(N-N_{\vec{\mbf{0}}},T)\ .
\label{state}
\end{equation}

Here, each Fock state $\ket{N_{\vec{\mbf{0}}}}\bra{N_{\vec{\mbf{0}}}}$ is weighted with the 
condensate particle number distribution $p_{N}(N_{\vec{\mbf{0}}},t)$, simultaneously capturing the 
distribution of non-condensate particles, as the particle number in the gas is conserved. 
Each state $\hat{\rho}_{\perp}(N-N_{\vec{\mbf{0}}},T)$ in eq.~(\ref{state})
denotes a thermal state projected onto the subspace of $(N-N_{\vec{\mbf{0}}})$ non-condensate particles, 
given that $N_{\vec{\mbf{0}}}$ particles populate the condensate mode:

\begin{equation}
\hat{\rho}_{\perp}(N-N_{\vec{\mbf{0}}},T)=\frac{
\hat{\mathcal{Q}}_{N-N_{\vec{\mbf{0}}}}{\rm e}^{-\beta\hat{\mathcal{H}}_\perp}
\hat{\mathcal{Q}}_{N-N_{\vec{\mbf{0}}}}}{\mathcal{Z}_\perp(N-N_{\vec{\mbf{0}}},T)}\ ,
\label{thermal}
\end{equation}

with $\mathcal{Z}_\perp(N-N_{\vec{\mbf{0}}},T) = \sum_{\lbrace N_{k}\rbrace}
{\rm exp}[-\sum_{\vec{\mbf{k}}\ne0}\beta\epsilon_{\vec{\mbf{k}}}N_{\vec{\mbf{k}}}]$, 
the partition function of $(N-N_{\vec{\mbf{0}}})$ indistinguishable 
particles, in which the sum of the tuples $\lbrace N_{\vec{\mbf{k}}_1}, N_{\vec{\mbf{k}}_2}, \ldots\rbrace$ 
is taken such that the total atom number is conserved, $\sum_{\vec{\mbf{k}}}N_{\vec{\mbf{k}}}=(N-N_{\vec{\mbf{0}}})$. 
The operator $\hat{\mathcal{H}}_\perp=\sum_{\vec{\mbf{k}}\ne\vec{\mbf{0}}}\epsilon_{\vec{\mbf{k}}}\hat{a}^\dagger_{\vec{\mbf{k}}}
\hat{a}_{\vec{\mbf{k}}}$ denotes the Hamiltonian of the non-condensate thermal vapor, $\beta=(k_BT)^{-1}$ the inverse temperature of the gas, 
and $\hat{\mathcal{Q}}_{N-N_{\vec{\mbf{0}}}}$ is a projector onto the Fock space of 
non-condensate number states with $(N-N_{\vec{\mbf{0}}})$ particles.

The fact that the N-body state in eq. (\ref{thermal}) is not a product state of a 
condensate and non-condensate density matrix doesn't cause fundamental problems to derive 
a master equation for the reduced condensate density matrix. More explicit, as shown in Refs.~\cite{TBE, TBE1}, the state in eq.~(\ref{thermal}) 
allows for the derivation of the master equation for a three-dimensional harmonic trapping potential 
without further approximations. Since the $N$-body state is diagonal in Fock number represenation, we focus 
on the evolution of the diagonal elements characterized by
$p_{N}(N_{\vec{\mbf{0}}},t)=\bra{N_0}\hat{\rho}_{\vec{\mbf{0}}}(t)\ket{N_{\vec{\mbf{0}}}}=\bra{N_{\vec{\mbf{0}}}}
{\rm Tr}_{\perp}\hat{\rho}^{{\rm (N)}}(t)\ket{N_{\vec{\mbf{0}}}}$ in eq.~(\ref{state}).
The corresponding master equation for the condensate 
number distribution in a gas of exactly $N$ atoms describes Bose-Einstein condensation as due 
to quantum jumps $N_{\vec{\mbf{0}}}\rightarrow N_{\vec{\mbf{0}}}\pm1$ 
of the condensate particle number: 

\begin{equation}
\begin{split}
\frac{\partial p_{N}(N_{\vec{\mbf{0}}},t)}{\partial t} =&- \left[\xi^{+}_{N}\left(N_{\vec{\mbf{0}}},T\right)+
\xi^{-}_{N}(N_{\vec{\mbf{0}}},T)\right]p_{N}(N_{\vec{\mbf{0}}},t) \\
&+ \xi^{+}_{N}(N_{\vec{\mbf{0}}}-1,T)p_{N}(N_{\vec{\mbf{0}}}-1,t) \\
& + \xi^{-}_{N}(N_{\vec{\mbf{0}}}+1,T)p_{N}(N_{\vec{\mbf{0}}}+1,t) \ ,  
\end{split}
\label{prop_evolution}
\end{equation}
with a condensate feeding rate $\xi^{+}_{N}(N_{\vec{\mbf{0}}},T)=2(N_{\vec{\mbf{0}}}+1)\lambda_{\rightsquigarrow}^{+}(N-N_{\vec{\mbf{0}}},T)$, 
and a condensate loss rate $\xi^{-}_{N}(N_{\vec{\mbf{0}}},T)=2N_{\vec{\mbf{0}}}\lambda_{\rightsquigarrow}^{-}(N-N_{\vec{\mbf{0}}},T)$. 
The single-particle transition rates
$\lambda_{\rightsquigarrow}^{\pm}(N-N_{\vec{\mbf{0}}},T)$ are given in Eq.~(\ref{rate1}). 

Remarkably, even though eq.~(\ref{prop_evolution}) describes Bose-Einstein condensation realistically 
in terms of two-body atomic collisions~\cite{Stoof, Walser1, TBE}, it \textit{formally} obeys 
the master equation for an ideal gas coupled to a thermal reservoir~\cite{Scully2}.
In contrast to the non-interacting case, however, the master eq.~(\ref{prop_evolution}) accounts for 
the real-time dynamics of Bose-Einstein condensation in a dilute, weakly interacting gas.
The  master equation (\ref{prop_evolution}) is valid up to order $a\varrho^{1/3}\ll1$. 
Thus, even though we neglect terms $\mathcal{O}(a\varrho^{1/3})$ for the sake of simplicity, the time 
scale for condensate formation depends on both the 
type \textit{and} the strength of the atomic two-body interactions. 
The condensation time is quantified by the single-particle transition rates 
$\lambda^{\pm}(N-N_{\vec{\mbf{0}}},T)$ in eq.~(\ref{prop_evolution}):

\begin{equation}
\begin{split}
\lambda_{\pm}(N_\perp,T)=&\frac{16\pi^{3}
\hbar^{2}a^{2}}{m^{2}}\sum_{\vec{\mbf{k}},\vec{\mbf{l}},\vec{\mbf{m}}\ne0}f_{\pm}
(\vec{\mbf{k}},\vec{\mbf{l}},\vec{\mbf{m}},N_{\perp},T)\times\\
&\times\delta^{{\rm (\Gamma)}}\left((\vec{\mbf{k}}+\vec{\mbf{l}}
-\vec{\mbf{m}})\cdot\vec{\mbf{\omega}}\right)\ ,
\end{split}
\label{rate1}
\end{equation}
with $f_{+}(\vec{\mbf{k}},\vec{\mbf{l}},\vec{\mbf{m}},N_{\perp},T)$ given by
\begin{equation}
\begin{split}
f_{+}(\vec{\mbf{k}},\vec{\mbf{l}},\vec{\mbf{m}},N_{\perp},T) 
&= \langle N_{\vec{\mbf{k}}}\rangle(N_\perp,T)\langle N_{\vec{\mbf{l}}}\rangle(N_\perp,T)\times\\
&\times\left[\langle N_{\vec{\mbf{m}}}\rangle(N_\perp,T)+1\right]
\vert\zeta_{\vec{\mbf{k}}\vec{\mbf{l}}}^{\vec{\mbf{m}}\vec{\mbf{0}}}\vert^{2}\ ,
\end{split}
\label{rate2}
\end{equation}
describing feeding processes of 
the condensate by one non-condensate particle, and with 
$f_{-}(\vec{\mbf{k}},\vec{\mbf{l}},\vec{\mbf{m}},N_{\perp},T)$ given by
\begin{equation}
\begin{split}
f_{-}(\vec{\mbf{k}},&\vec{\mbf{l}},\vec{\mbf{m}},N_{\perp},T) = 
\left[\langle N_{k}\rangle(N_{\perp},T)+1\right]\times\\
&\times\left[\langle N_{l}\rangle(N_{\perp},T)+1\right] 
\langle N_{\vec{\mbf{m}}}\rangle(N_{\perp},T)\vert\zeta_{\vec{\mbf{k}}\vec{\mbf{l}}}^{\vec{\mbf{m}}\vec{\mbf{0}}}\vert^{2}\ ,
\end{split}
\label{rate3}
\end{equation}
which quantifies losses of condensate particles 
towards the non-condensate single-particle modes originating from atomic interactions. 

Condensate feedings and losses with microscopic 
energy balances $\epsilon_{\vec{\mbf{k}}}+\epsilon_{\vec{\mbf{l}}}=\epsilon_{\vec{\mbf{m}}} + \epsilon_{\vec{\mbf{0}}}\pm\hbar\Delta$ 
are quantified in eqs.~(\ref{rate2},~\ref{rate3}) proportional to the terms 
$\zeta_{\vec{\mbf{k}}\vec{\mbf{l}}}^{\vec{\mbf{m}}\vec{\mbf{0}}}=\int{\rm d}\vec{\mbf{r}}~
\Psi_{\vec{\mbf{k}}}(\vec{\mbf{r}})\Psi_{\vec{\mbf{l}}}(\vec{\mbf{r}})\Psi^{\star}_{\vec{\mbf{m}}}(\vec{\mbf{r}})\Psi^\star_{\vec{\mbf{0}}}(\vec{\mbf{r}})$, 
where  $\epsilon_{\vec{\mbf{k}}}=\hbar\vec{\mbf{k}}\vec{\mbf{\omega}} + \hbar(\omega_x+\omega_y+\omega_z)/2$ with 
$\vec{\mbf{k}}=(k_x,k_y,k_z)$ are single-particle energies,
and where $\Delta$ is the off-resonance of a given two-body process and $\Gamma$ the energy uncertainty, in analogy 
to the quantum optical case \cite{Tan}. Thus, the $\delta$-distribution in eq.~(\ref{rate1}) reflects 
and ensures energy conservation on a certain width $\Gamma$ 
arising from the \textit{finite} decay time of non-condensate correlations \cite{TBE, TBE1}.
The $\langle N_{\vec{\mbf{k}}}\rangle(N_{\perp},T)$ denote single-particle occupations, 
which are calculated with respect to each non-condensate thermal mixture 
of $N_\perp=(N-N_{\vec{\mbf{0}}})$ particles in eq.~(\ref{thermal}), 
obeying $\sum_{k}\langle N_{k}\rangle(N_{\perp},T)=(N-N_{\vec{\mbf{0}}})$. 
The evaluation of $\Gamma$ is beyond the scope of the present article, and wil be presented elsewhere \cite{TBE}. 

\begin{figure}[t] 
\psfrag{xlabel}{$t$}
\psfrag{zlabel}{$\mbf{p(N_{\vec{\mbf{0}}},t)}$}
\psfrag{ylabel}{$\mbf{N_{\vec{\mbf{0}}}}$}
\psfrag{xlabel1}{\begin{small}\textbf{time}\end{small} $\mbf{t}$ \begin{small}\textbf{[s]}\end{small}}
\psfrag{ylabel1}{$\mbf{\sigma_{0}(t)}$}
\psfrag{xlabel2}{\begin{small}\textbf{time}\end{small} $\mbf{t}$ \begin{small}\textbf{[s]}\end{small}}
\psfrag{ylabel2}{$\mbf{\sqrt{\Delta^{2} N_{\vec{\mbf{0}}}(t)} {\rm~}[\sqrt{N}]}$}
\centerline{\includegraphics[width=6.2cm, height=5.2cm, angle = 0.0]{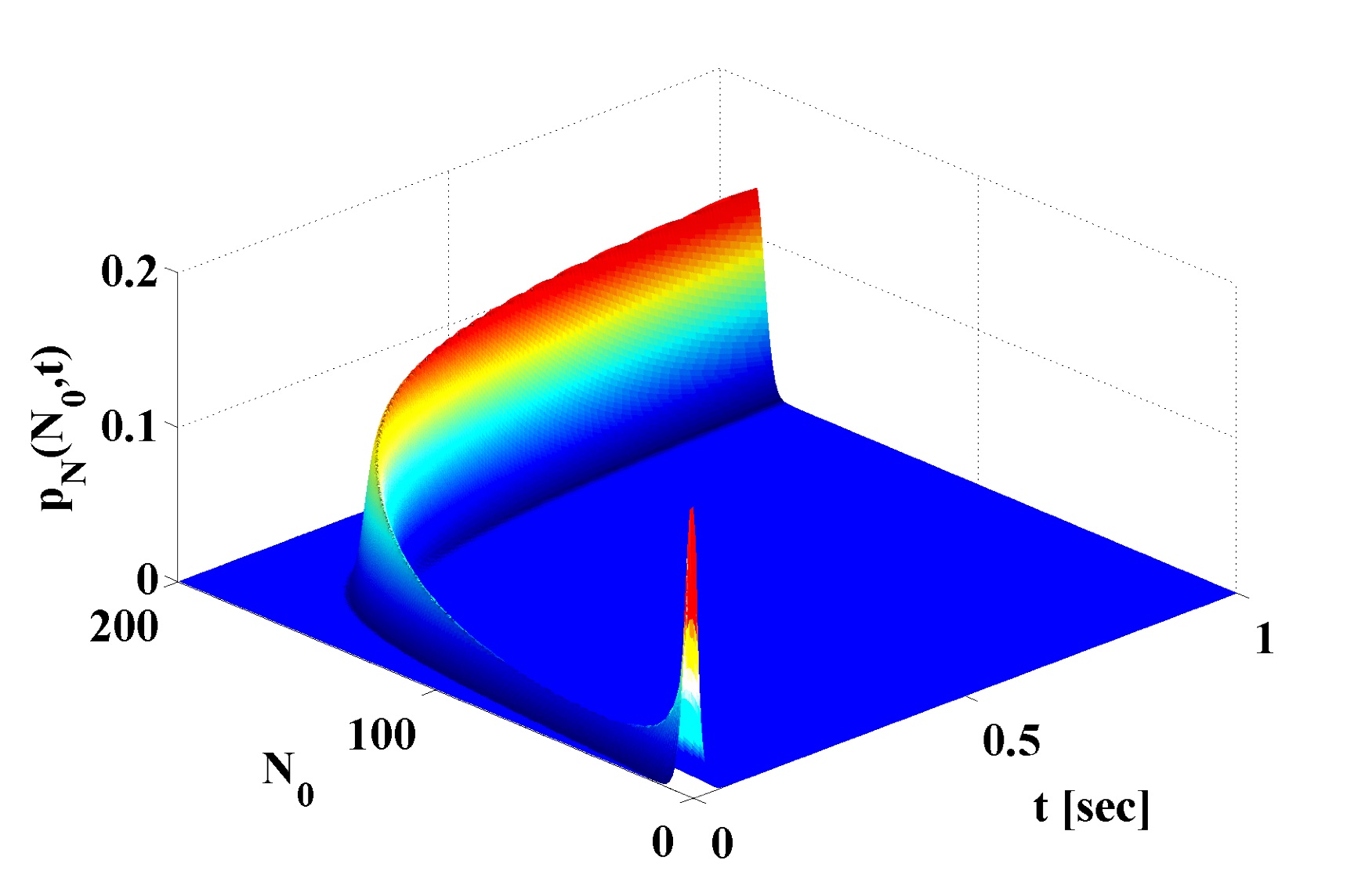}\vspace{0.1cm}}
\caption{(color online) Evolution of condensate particle number distribution $p_{N}(N_{\vec{\mbf{0}}},t)$ 
during the transition of a gas of $N=200$ ${\rm Rb}^{87}$ interacting atoms
into a Bose-Einstein condensate of final temperature $T/T_c=0.4$, obtained from eq.~(\ref{prop_evolution}).}
\label{fig1}
\end{figure}

\section{III. Dynamics of Bose-Einstein condensation}
\begin{figure}[t]
\psfrag{xlabel1}{\begin{small}\textbf{time}\end{small} $\mbf{t}$ \begin{small}\textbf{[s]}\end{small}}
\psfrag{ylabel1}{$\mbf{\sigma_{\vec{\mbf{0}}}}(t)$}
\psfrag{xlabel2}{\begin{small}\textbf{time}\end{small} $\mbf{t}$ \begin{small}\textbf{[s]}\end{small}}
\psfrag{ylabel2}{$\mbf{\sqrt{\Delta^{2} N_{\vec{\mbf{0}}}(t)} {\rm~}[\sqrt{N}]}$}
\psfrag{label}{$\mbf{t = 125}$ \begin{small}\textbf{ms}\end{small}}
\psfrag{label2}{$\mbf{\sigma_{\vec{\mbf{0}}}} = 0.25$}
\centerline{\includegraphics[width=6.8cm, height=4.0cm, angle = 0.0]{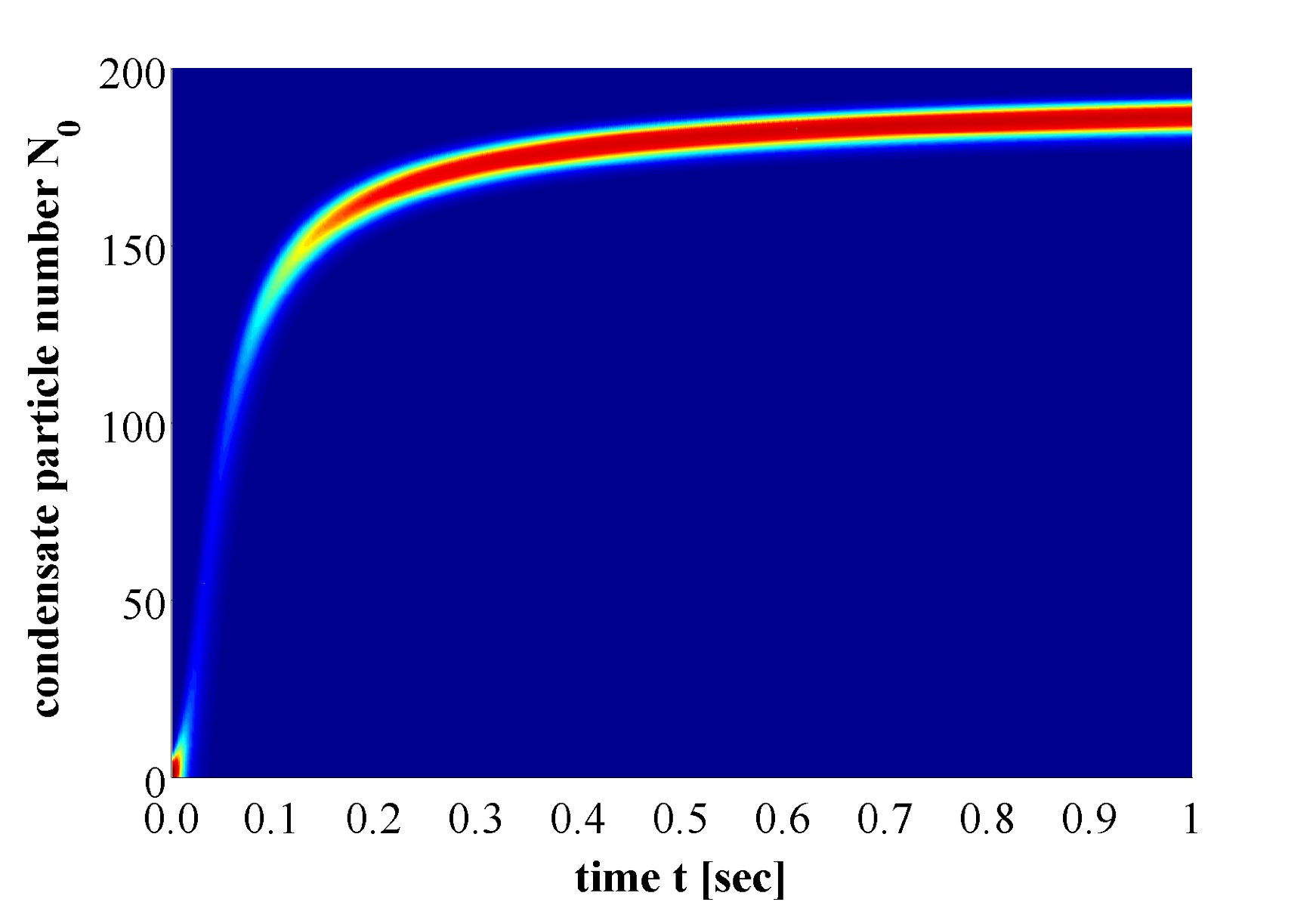}\hspace{0.5cm}}
\centerline{\includegraphics[width=3.6cm, height=2.8cm, angle = 0.0]{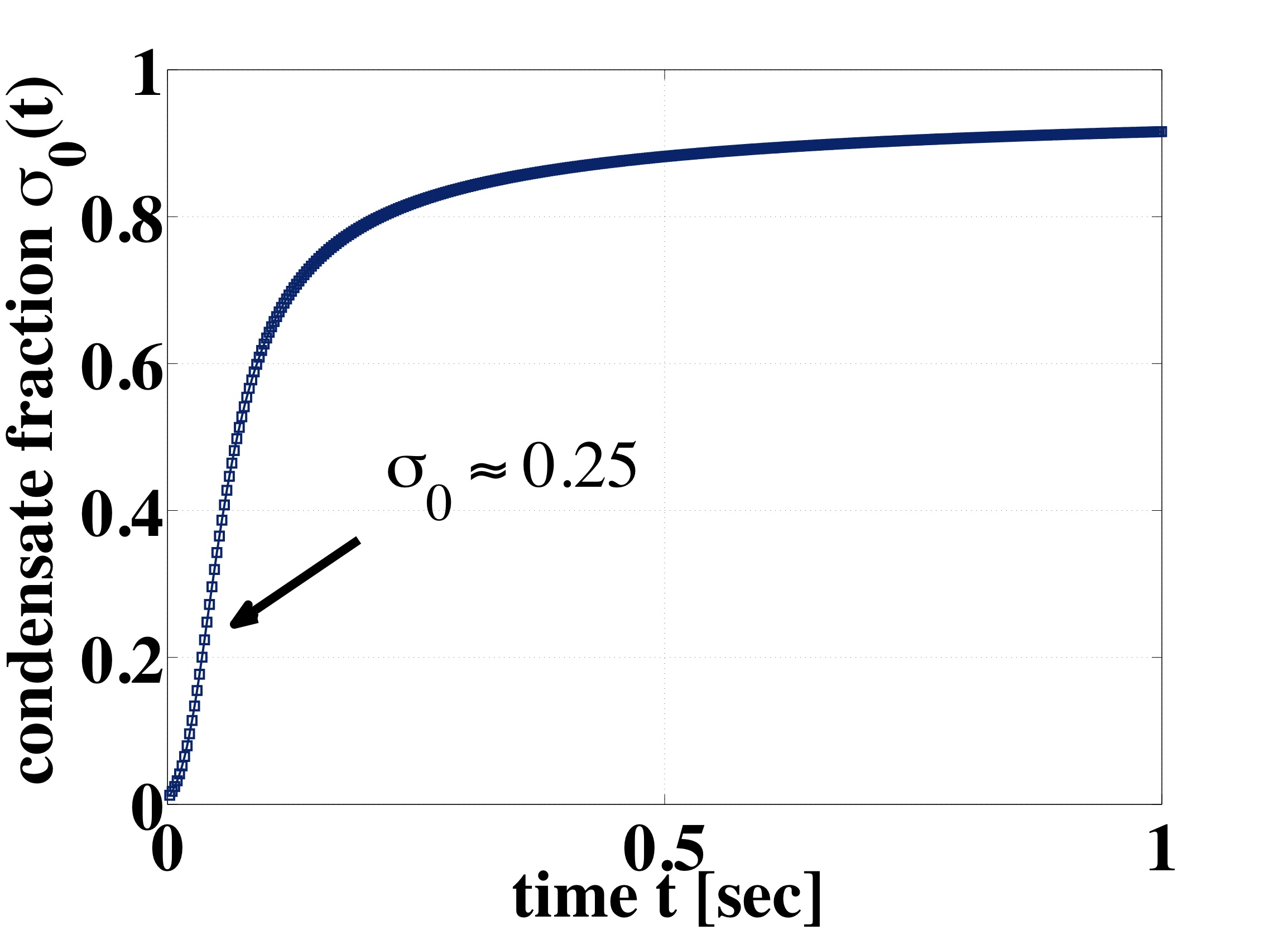}\hspace{0.5cm}}
\centerline{\includegraphics[width=3.6cm, height=2.8cm, angle = 0.0]{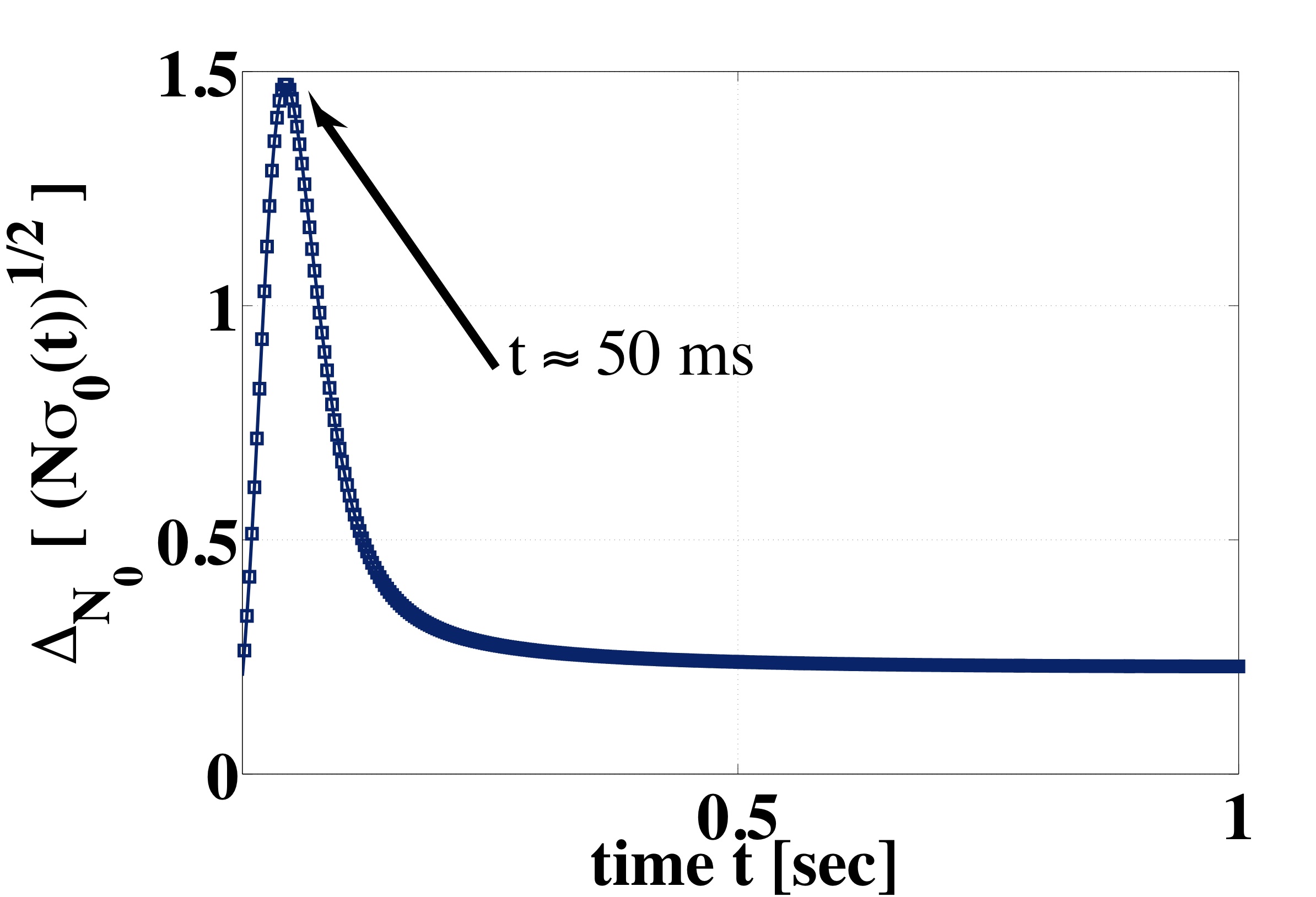}}
\caption{(color online) X-y projection of the distribution $p_N(N_{\vec{\mbf{0}}},t)$ from fig.~\ref{fig1} (upper panel). 
Average condensate fraction $\sigma_0(t)$ (lower left panel) and 
width $\Delta_{N_0}(t)$ in units $(N\sigma_0(t))^{1/2}$ of the distribution $p_{N}(N_{\vec{\mbf{0}}},t)$ 
(lower right panel).}
\label{fig2}
\end{figure}

The dynamics of the condensate number distribution during Bose-Einstein condensation 
is displayed in figs.~\ref{fig1} and \ref{fig2} as a function of time $t$ and condensate particle number $N_{\vec{\mbf{0}}}$, after a sudden 
switch of the atomic cloud's temperature $T$ below the 
ideal gas critical temperature $T_c$ expected for Bose-Einstein condensation~\cite{condensation}.
For numerical propagation of eq.~(\ref{prop_evolution}), we consider a small Bose gas of $N=200$ 
weakly interacting $^{87}{\rm Rb}$ atoms (with s-wave scattering length $a=5.4$ nm~\cite{SWAVE, Stringari}), 
prepared in a time-independent, three-dimensional harmonic trapping potential with trapping frequencies 
$\vec{\mbf{\omega}}=(\omega_x=2\pi\times42.0~{\rm Hz}, \omega_y=2\pi\times42.0~{\rm Hz}, \omega_z=2\pi\times120.0~{\rm Hz})$.
Despite the well-known S-shape behavior \cite{QKT, Walser1} of the average condensate occupation number, the condensate number distribution in particular gives rise of non-condensate number fluctuations during the condensation process \cite{Obe1}, since the total number of particles in the gas is fixed. Characteristic for a phase transition, we observe an initial spread of the condensate number distribution (large number fluctuations in the thermal vapor). While the condensate grows, these incipiently large fluctuations 
reduce until the reaching of a steady state. This steady state is in particular 
characterized in that the \textit{net} particle flow between condensate and 
non-condensate is zero. 

Achieving Bose-Einstein condensation hence relies on the atomic 
interactions in order to bring the gas into equilibrium in the quantum degenerate regime, 
number fluctuations are prominent during condensate formation 
as highlighted by figs.~\ref{fig1} and \ref{fig2}, and energy uncertainty (accounted for by the width $\Gamma$ of the delta function in eq. (\ref{rate1})) 
is present due to the finite decorrelation time between condensate and non-condensate particles. 
Can it thus really be that -- according to the fundamental laws of thermodynamics -- 
a mesoscopic, weakly interacting Bose-Einstein condensate obeys the 
Gibbs-Boltzmann statistics?

\section{IV. Equilibrium state}
To answer this question, let us recall that the steady state solution of the condensate number distribution in 
eq.~(\ref{prop_evolution}) entirely defines the equilibrium steady state of gas in eq.~(\ref{state}).
In our framework, the key aspect that a Bose-Einstein condensate obeys 
the Boltzmann statistics lies in that quantum coherences in the 
gas rapidly decay on a certain time scale $\Gamma^{-1}$ and in that the gas is sufficiently dilute $a\varrho^{1/3}\ll1$.  
For sufficiently small energy uncertainties corresponding to the formal limit 
$\beta\hbar\Gamma\rightarrow0^+$, i.e. sufficiently fast non-condensate equilibration, 
$\beta\hbar\Gamma\ll1$, the rates $\lambda^{+}(N-N_{\vec{\mbf{0}}},T)$ and $\lambda^{-}(N-N_{\vec{\mbf{0}}},T)$ obey 
the balance condition $\lambda^{+}(N-N_{\vec{\mbf{0}}},T)= 
z(N-N_{\vec{\mbf{0}}},T)\lambda^{-}(N-N_{\vec{\mbf{0}}},T)$, where 

\begin{equation}
z(N-N_{\vec{\mbf{0}}},T) = {\rm e}^{-\beta\epsilon_{\vec{\mbf{0}}}}
\frac{\mathcal{Z}_\perp(N-N_{\vec{\mbf{0}}}-1,T)}{\mathcal{Z}_\perp(N-N_{\vec{\mbf{0}}},T)}\ .
\label{final2}
\end{equation}

In this case, the unique steady solution of eq.~(\ref{prop_evolution}) is determined by 
$p_N(N_{\vec{\mbf{0}}},T)\sim\prod_{k=1}^{N_{\vec{\mbf{0}}}-1}\xi^{+}_{N}(k-1,T)
/\xi^{-}_{N}(k,T)$. Using that $\lambda^-(N-N_{\vec{\mbf{0}}})\simeq\lambda^-(N-N_{\vec{\mbf{0}}}+1)$, 
this steady state can be related to the balance condition in eq.~(\ref{final2}), which leads to:

\begin{equation}
p_N(N_{\vec{\mbf{0}}},T)={\rm e}^{-\beta \epsilon_{\vec{\mbf{0}}}N_{\vec{\mbf{0}}}}
\frac{\mathcal{Z}_\perp(N-N_{\vec{\mbf{0}}},T)}{\mathcal{Z}(N,T)}\ ,
\label{final}
\end{equation}

where $\mathcal{Z}_\perp(N-N_{\vec{\mbf{0}}},T)$ is the partition function of the non-condensate gas in 
eq.~(\ref{thermal}), and where $\mathcal{Z}(N,T)$ is the number of accessible microstates 
of the entire system, condensate and non-condensate.
From eq.~(\ref{final}), it thus follows that the condensate statistics is independent on the specific strength 
of the two-body collisions once the steady state has been reached. 
This result is valid in the limit of dilute atomic gases and weak interactions, $a\varrho^{1/3}\ll1$, as has been also verified 
numerically (innluding the finite with $\Gamma$ of condensate feeding and loss rates). According to eqs.~(\ref{state}) and~(\ref{final}), the steady $N$-body state reached by the atomic collisions remarkably thus obeys the statistics of a thermal Boltzmann state of $N$ bosonic, non-interacting particles~\cite{Scully2}:

\begin{equation}
\hat{\sigma}^{{\rm (N)}}(t\rightarrow\infty) = \hat{\mathcal{Q}}_N\frac{{\rm e}^{-\beta\hat{\mathcal{H}}}}{\mathcal{Z}(N,T)}\hat{\mathcal{Q}}_N + \mathcal{O}(a\varrho^{1/3})\ , 
\label{thermal2}
\end{equation} 

with $\hat{\mathcal{H}}=\sum_{\vec{\mbf{k}}}\epsilon_{\vec{\mbf{k}}}
\hat{a}^\dagger_{\vec{\mbf{k}}}\hat{a}_{\vec{\mbf{k}}}$, 
the Hamiltonian of a non-interacting gas, $\mathcal{Z}(N,T)$ the partition function of 
$N$ indistinguishable particles, and $\hat{\mathcal{Q}}_N$, the projector onto the Fock subspace 
of $N$ particles. 

\section{V. Conclusion}
In conclusion, we have monitored the condensate and non-condensate 
number distribution during the phase transition of a mesoscopic Bose gas into 
a Bose-Einstein condensate. We find that the steady
state of the gas in the condensed phase is a 
Gibbs-Boltzmann thermal state, obeying the Bose-Einstein statistics 
of an ideal gas for sufficiently dilute and weakly interacting atomic gases. 
The reported scenarios were numerically 
reproducable for total particle numbers up to $N=10^5$~\cite{TBE1}. 
The impact of environmental decoherence sources 
onto the condensate formation process during condensate formation such as 
detailed study of deviations of the final steady state from the Gibbs-Boltzmann 
equilibrium will be of future interest and discussed elsewhere. 

\acknowledgments
I thank Andreas Buchleitner, Dominique Delande, Boris Fine, Cord M\"{u}ller, Alice Sinatra and Thomas Wellens 
for helpful discussions and comments during the development of the theory. Financial support 
from Heidelberg University for the editing of this theory \cite{TBE1} is acknowledged.

\end{document}